\documentclass[aps,twocolumn]{revtex4}
\usepackage{epsfig}
\newcommand{\be}{\begin{equation}}
\newcommand{\ee}{\end{equation}}
\newcommand{\ba}{\begin{eqnarray}}
\newcommand{\ea}{\end{eqnarray}}

\begin{document}

\title{Tau neutrino propagation and tau energy loss}
\author{Sharada Iyer Dutta}
\affiliation{%
Department of Physics and Astronomy, 
State University of New York at Stony Brook,\\
Stony Brook, New York 11794
USA}
\author{Yiwen Huang and Mary Hall Reno}
\affiliation{%
Department of Physics and Astronomy, University of Iowa,
Iowa City, Iowa 52242 USA}
\begin{abstract}
Electromagnetic energy loss of tau leptons is an important ingredient
for eventual tau neutrino detection from high energy astrophysical
sources. Proposals have been made to use mountains as neutrino
converters, in which the emerging tau decays in an air shower.
We use a stochastic evaluation of both tau neutrino conversion
to taus and of tau electromagnetic energy loss. We examine the
effects of the propagation for mono-energetic incident tau neutrinos
as well as for several neutrino power-law spectra.
Our main result is a parameterization of the tau electromagnetic
energy loss parameter $\beta$.  We compare the results from the
analytic expression for the tau flux using this $\beta$
with other parameterizations of $\beta$.

\end{abstract}

\maketitle

\section{Introduction}

The observed neutrino fluxes in solar neutrino experiments \cite{solar}
and in atmospheric neutrino experiments \cite{atm} lead to a picture of
neutrino oscillations with substantial $\nu_\mu\rightarrow \nu_\tau$
mixing \cite{mix}.
For high energies over astronomical distances, astrophysical
sources of $\nu_e$ and $\nu_\mu$ can become
sources of $\nu_\tau$ as well. One method of detecting $\nu_\tau$
from astrophysical sources is to look for the charged particles
they produced in or near neutrino telescopes.

Because of the short lifetime of
the tau, $\nu_\tau\rightarrow \tau\rightarrow\nu_\tau$ regeneration
can be an important effect as the $\nu_\tau$
passes through a significant column depth through the Earth \cite{halzen,
regeneration,bottai,bmss,fargionboth}.  
This is typically relevant for neutrino energies below
$E_\nu\sim 10^8$ GeV. At higher energies, one can imagine using the Earth as
a tau neutrino converter 
\cite{fargionboth,yoshida,reya,feng,aramo,convertfar,bertou,
athar,jones,yeh,cao,huang}. 
Because of the relatively
short interaction length of the neutrino at high energies,
column depths on the order of the distance through a mountain
or from small skimming angles relative to the horizon are most
important.
Following the tau neutrino conversion,
a tau is detected by its decay e.g., in ice with IceCube, 
RICE or the proposed
ANITA detectors, or after it emerges into the air.
The possibility of detectable signals depends on the incident fluxes,
of which there are a variety of models \cite{models}, and on
the detectors and their sensitivity to various energy regimes.

Theoretical calculations often rely on approximating the neutrino interaction
and tau energy loss in terms of analytic functions of energy
and column depth \cite{feng,aramo}. These analytic tools
are important in exploring the possibility for tau neutrino
signals in a variety of detectors, even though ultimately a full
simulation of tau production and energy loss is required.  
In the energy regime where $E_\nu\gg 10^8$ GeV Huang, Tseng and Lin
have made an evaluation of the smearing of energy due to 
tau propagation by doing an approximate stochastic evaluation of
electromagnetic energy loss of the tau \cite{huang}. Looking at fixed
incident neutrino energies, they find that the relative energy 
fluctuation of the tau can become large at high energies, especially
for larger column depths. This has the potential to make 
interpretation of fluxes difficult at high energies.

In this paper, we evaluate tau neutrino and tau  propagation 
using a stochastic evaluation of tau electromagnetic energy loss based on the
work of Ref. \cite{drss}. We find the approximate evaluation 
of Ref. \cite{huang} reliable. In addition to considering
mono-energetic incident neutrinos, we also consider incident neutrino
fluxes with a power law behavior, $E_\nu^{-\gamma}$. We compare
the flux of taus following propagation through 10 km rock and
100 km rock using the stochastic evaluation with estimates
of the emerging flux of taus using analytic
approximations. We refine some recent parameterizations
of the tau energy loss as a function of tau energy\cite{feng,aramo}
to better approximate the stochastic result for the flux
of emerging taus  using 
analytic approximations. The best parameterization of the electromagnetic
energy loss parameter $\beta$ is with a logarithmic dependence on 
tau energy. 

In the next section, we describe the models for the high
energy neutrino nucleon cross section and the tau energy loss used in
our Monte Carlo simulation. Results of the simulation and analytic
approximations for the tau energy as a function of distance traveled,
for monoenergetic neutrinos, appear in Section III. Section IV
includes results and analytic approximations for power law 
incident neutrino
spectra. A summary of our results and a comparison with a simplified
analytic approach are in
the final section.

\section{Neutrino interactions and tau energy loss}

Our evaluation of neutrino interactions and tau energy loss is
performed using a one-dimensional Monte Carlo computer simulation.
The neutrino charged current and neutral current interactions are
calculated based on neutrino-isoscalar nucleon cross sections
\cite{gqrs95,gqrs98} using the
CTEQ6 parton distribution functions (PDFs)\cite{cteq6}. 

At ultra-high energies, there is a competing effect in the neutrino-nucleon
cross section
as a function of $Q^2=-q^2>0$, the quantity describing the four-momentum
squared of the vector boson.  PDFs increase 
as a function of $\ln Q^2$ \cite{andreev}, 
however,
the vector boson propagator effectively
cuts off the growth in $Q^2$ at the vector boson mass. 
For 
charged current scattering, this means that the value of
parton $x\equiv Q^2/2p\cdot q$ for nucleon four-momentum $p$ is
approximately
\begin{equation}
x\sim \frac{M_W^2}{ME_\nu}
\end{equation}
in terms of the incident neutrino energy $E_\nu$, nucleon mass
$M$ and $W$ boson mass $M_W$. For $E_\nu=10^{12}$ GeV, one finds that
$x\sim 10^{-8}$. 

\begin{figure}[t]
\begin{center}
\epsfig{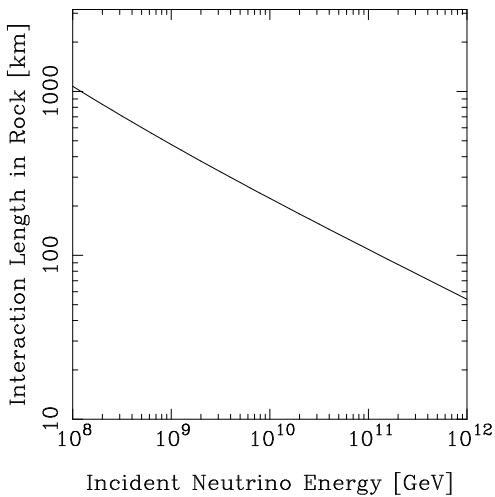}
\end{center}
\caption{Neutrino interaction length $L_{\rm int}$ in rock as a function of
neutrino energy.}
\label{fig:lint}
\end{figure}

The CTEQ6 PDFs are parameterized for
$10^{-6}<x<1$, so one must extrapolate for smaller values of $x$.
The
PDFs at small values of parton $x$ ($x<10^{-6}$) are extrapolated
here using a power law $xq(x,Q^2)\sim x^{-\lambda}$
matched to DGLAP evolved PDFs
at $x_{min}=10^{-6}$:
\begin{equation}
xq(x_{min},M_W^2)=A x_{min}^{-\lambda}\ .
\end{equation}
This approach is based on the result \cite{ekl} that for $\lambda\sim 0.3-0.4$,
the gluon PDF has the approximate form of $xg(x,Q)\sim x^{-\lambda}$,
and $g\rightarrow q\bar{q}$ splitting is responsible for the sea
quark distributions that dominate the cross section at ultra-high
energies.
We show in Fig. \ref{fig:lint} the neutrino
interaction
length in rock. Other approaches to the small $x$ extrapolations
for ultra-high energy neutrino cross sections yield similar interaction
lengths \cite{others}, to within a factor of approximately
$2^{\pm 1}$ at $10^{12}$ GeV.

The differential distribution for neutrino
nucleon scattering in terms of the lepton inelasticity
$y$,
\begin{equation}
y\equiv \frac{E_\nu - E_\tau^i}{E_\nu}\ ,
\end{equation}
is evaluated similarly. We indicate the energy of the tau at the
point of its production by $E_\tau^i$.
The neutrino interaction probability and
resulting
tau energy is evaluated using a one-dimensional Monte Carlo program.
In our analytic approximations described below, we use $\langle
y\rangle
\simeq 0.2$, which is approximately correct at ultra-high energies
\cite{gqrs95}.
An indication of this is shown by the solid line in
Fig. \ref{fig:eavg},
which shows the ratio of the average tau energy to the initial
neutrino
energy, as a function of neutrino energy, in the absence of
electromagnetic
energy loss corrections.

Our evaluation of the tau electromagnetic energy loss follows the
procedure described in Ref. \cite{drss}. Based on the
scheme outlined by Lipari and Stanev \cite{ls} and others
\cite{music}, we have incorporated the electromagnetic mechanisms
of ionization, bremsstrahlung (brem), $e^+e^-$ pair production
(pair) and photonuclear (nuc) scattering. The tau lifetime
is taken into account.

The average energy lost per unit distance is commonly described
by the formula
\begin{equation}
\label{eq:dedx}
\langle \frac{dE_\tau}{dz}\rangle = -(\alpha +\beta E_\tau)\rho\ .
\end{equation}
In this expression, $z$ is the distance the tau travels. The quantity
$\alpha \simeq 2\times 10^{-3}$ GeV cm$^2$/g accounts for ionization
energy loss \cite{ionization}. Here, $\beta$ is
\begin{eqnarray}
\beta &=& \beta^{\rm brem}+\beta^{\rm pair}+\beta^{\rm nuc} \nonumber \\
&=&\sum_i\frac{N_A}{A} \int dy\, y\, \frac{d\sigma^i}{dy}\ ,
\end{eqnarray}
for Avogadro's number $N_A$, atomic mass number $A$, tau inelasticity
$y$ and $i={\rm brem,\ pair}$ and nuc.

In our treatment here, as in Ref.
\cite{drss} we separate $\beta$ into continuous and
stochastic terms:
\begin{eqnarray}
\beta & = & \beta_{cont}+\beta_{stoc} \nonumber \\
&=&\frac{N_A}{A}\int_0^{10^{-3}}dy\, y\,\frac{d\sigma^i}{dy} + 
\frac{N_A}{A}\int_{10^{-3}}^1dy\, y\,\frac{d\sigma^i}{dy}\ .
\label{eq:split}
\end{eqnarray}
We treat the tau propagation stochastically, with
a series of electromagnetic interactions occurring according to
probabilities based on $d\sigma^i/dy$ for $y>10^{-3}$ and
the decay probability. For
each distance between interactions
in the Monte Carlo program, the tau energy loss calculated
from Eq. (3) is supplemented by  Eq. (\ref{eq:dedx}) with the
substitution of $\beta\rightarrow \beta_{cont}$.
By doing it this way, we account for the full electromagnetic
energy loss, but we limit computer time spent where the cross section
is large, but the energy loss is small ($y<10^{-3}$).
The detailed formulas for the bremsstrahlung 
\cite{brem}, pair production \cite{pair} and photonuclear \cite{nuc}
processes are discussed and collected in Ref. \cite{drss}. See also
Ref. \cite{lkv} for the muon case.

For the muon, $\beta$ is dominated by pair production and bremsstrahlung
for $E_\mu=10^3-10^7$ GeV. At higher energies, $\beta $ grows slowly
with energy due to the photonuclear process which contributes
a larger fraction for $E_\mu>10^7$ GeV. 
For the tau, the photonuclear process is comparable to 
pair production, while the
bremsstrahlung contribution is suppressed, leading to an increase in 
$\beta$ with energy for $E_\tau>10^3$ GeV.

In the analytic approximations discussed below, we will consider several
forms of $\beta$ as a function of $E_\tau$, and we will treat
\begin{equation}
\label{eq:dedxapprox}
\langle \frac{dE_\tau}{dz}\rangle \simeq \frac{dE_\tau}{dz}\ .
\end{equation}
To guide our discussion, it is useful to compare the relevant constants
and distance scales. First, at $10^8$ GeV,
\begin{equation}
\beta E_\tau\simeq 60 \frac{{\rm cm}^2 {\rm GeV}}{{\rm g}}\Biggl(\frac{E_\tau}{10^8
\ {\rm GeV}}\Biggr)
\end{equation}
is much larger than $\alpha$, so we can essentially ignore $\alpha$.
The tau range is then characterized roughly
by
\begin{equation}
\frac{1}{\beta\rho}\sim 6\ {\rm km}
\end{equation}
in standard rock ($\rho = 2.65$ g/cm$^3$). In fact, the range
is somewhat larger than this estimate. The decay length of the
tau is 
\begin{equation}
\gamma c\tau\simeq 5\ {\rm km} \Biggl(\frac{E_\tau}{10^8
\ {\rm GeV}}\Biggr)\ .
\end{equation}
It is in the energy regime near $10^8$ GeV that the transition from
the lifetime dominated to the energy loss dominated range 
for the tau occurs.

\section{Mono-energetic neutrinos}

\subsection{Results from Monte Carlo simulation}

\begin{figure}[t]
\begin{center}
\epsfig{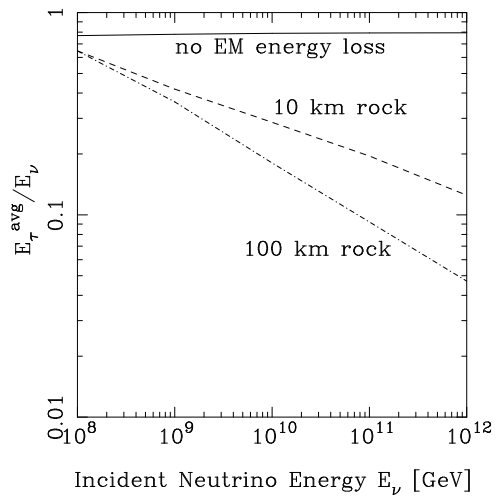}
\end{center}
\caption{Average tau energy scaled by the energy of incident neutrinos that produced the taus, without energy loss (solid), and with energy loss
for taus produced in 10 km and 100 km rock (dashed).}
\label{fig:eavg}
\end{figure}

\begin{figure}[t]
\begin{center}
\epsfig{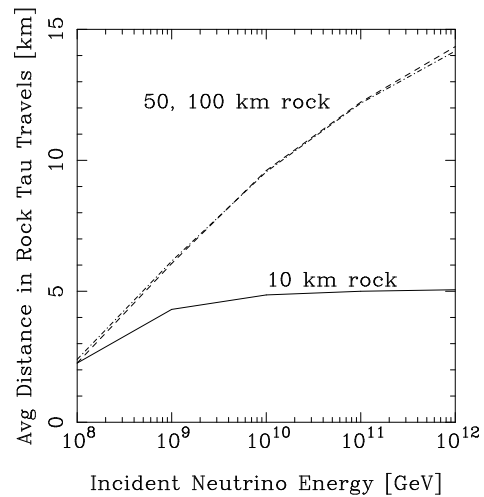}
\end{center}
\caption{Average distance the tau travels after production which
survives to emerge from 10 km rock (solid), 50 km rock (dashed) and
100 km rock (dot-dashed), as a function of incident neutrino energy.}
\label{fig:dleft}
\end{figure}

We begin our discussion by considering single incident 
neutrino energies
to see the effect of energy loss combined with decay probabilities
for two different trajectories: 
through 10 km rock (26.5 km.w.e.)
and through 100 km rock. Fig. \ref{fig:eavg}
shows the average tau energy divided by the
incident neutrino energy as a function of
incident
neutrino energy. The average distance that the tau travels after
production which survives to emerge from the rock (10, 50 and 100 km)
is shown in
Fig. \ref{fig:dleft}.

For 10 km of rock, above $10^9$ GeV, the average distance the tau
travels is 5 km. The neutrino interaction probability is equal 
anywhere in the 10 km path, and the energies are large enough that
the decay length is long compared to the path. For 50 km and 100 km,
Fig. \ref{fig:dleft} shows that the last $\sim 15$ km are the most important
for the emerging taus.

\begin{figure}[t]
\begin{center}
\epsfig{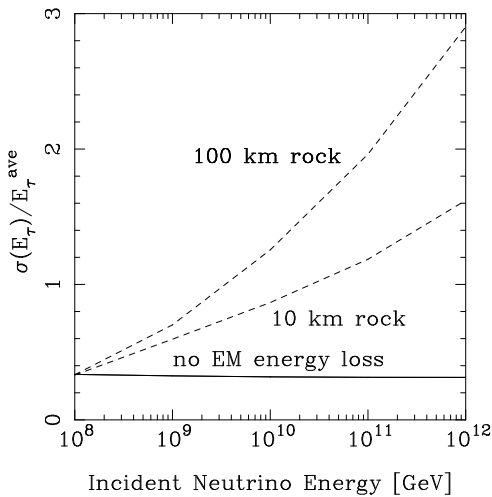}
\end{center}
\caption{Standard deviation $\sigma$ scaled by the average tau
energy as a function of the incident neutrino energy without
energy loss (solid) and for 10 km and 100 km of rock (dashed).}
\label{fig:sd}
\end{figure}

The standard deviation,
normalized by the average tau energy,
\begin{equation}
\frac{\sigma}{E_\tau^{avg}}= \frac{\sqrt{\langle E_\tau^2\rangle
-\langle E_\tau\rangle ^2}}{\langle E_\tau\rangle}
\label{eq:sd}
\end{equation}
is shown is Fig. \ref{fig:sd}. Higher energies and larger distances
lead to larger fluctuations. Huang et al.'s results in Ref. \cite{huang} correspond to our Figs. \ref{fig:eavg} and \ref{fig:sd} and are in good
agreement.
We comment that by normalizing the standard deviation to the incident
neutrino energy rather than average energy of the emerging tau, one
finds a more stable result for the ratio. The standard deviation scales
approximately as $(0.20-0.25)\times E_\nu$ for 10 km rock over the
same energy range as the figure, and with a factor of $(0.13-0.25)\times
E_\nu$ for 100 km rock.

\begin{figure}[t]
\begin{center}
\epsfig{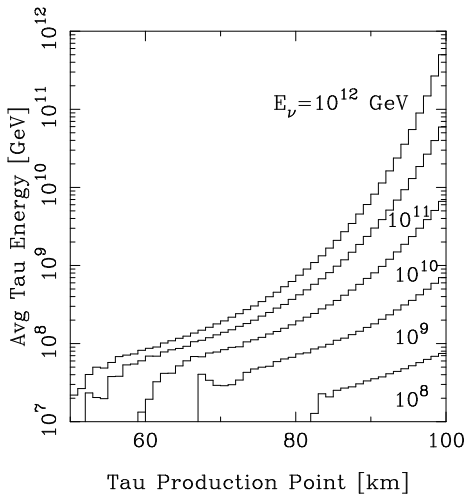}
\end{center}
\caption{Average energy of emerging tau as a function of 
its production point for a total distance of 100 km of rock, 
for fixed initial 
neutrino energies between $10^8-10^{12}$ GeV.}
\label{fig:evsz}
\end{figure}

A final result for mono-energetic neutrinos is shown in Fig. 
\ref{fig:evsz}. Each histogram shows the average energy of the emerging
tau as a function of the production point $z$ for a total depth of
100 km and at fixed initial neutrino energy. The energies of the
incident neutrinos are $E_\nu=10^n$ GeV for $n=8,9,...,12$ (lower to
upper curves).

\subsection{Analytic approximations}

\begin{figure}[t]
\begin{center}
\epsfig{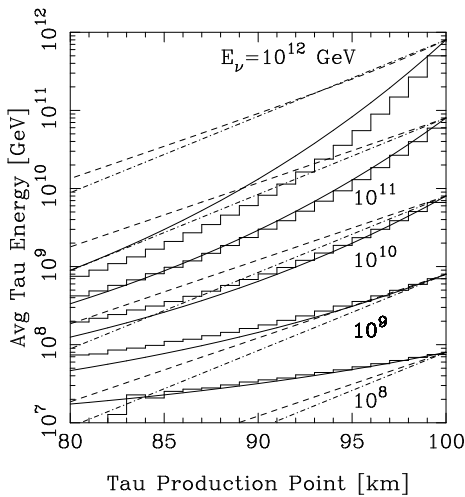}
\end{center}
\caption{Average energy of emerging tau as a function of 
its production point for a total distance of 100 km of rock, 
for fixed initial 
neutrino energies between $10^8-10^{12}$ GeV from Monte Carlo
simulation (histograms), and parameterization of $\beta$ as in
case (I) (dot-dashed), case (II) (dash) and case (III) (solid).}
\label{fig:evszcomp}
\end{figure}

Standard approaches to parameterizing the final energy
in terms of the initial energy begin with
\begin{equation}
\frac{dE_\tau}{dz}\simeq -\beta\rho E_\tau
\label{eq:dedx2}
\end{equation}
The solution depends on the energy dependence of $\beta$. Several
parameterizations have been used, including a linear
dependence on energy \cite{aramo} and a logarithmic energy
dependence \cite{fargionboth}. We consider three cases: 
\begin{equation}
\label{eq:cases}
\matrix{({\rm I}) & E_\tau=E_\tau^i e^{-\beta \rho z^\prime}\cr
& \cr
& \beta={\rm constant}\cr
 & \cr
({\rm II}) & E_\tau = \frac{E_\tau^i\beta_\tau e^{-\beta_\tau \rho z^\prime}}
{\beta_\tau +\gamma_\tau E_\tau^i (1- e^{-\beta_\tau \rho z^\prime})}\cr 
& \cr
& \beta= \beta_\tau + \gamma_\tau E_\tau \cr
& \cr
({\rm III}) & E_\tau = \exp\Bigl[-\frac{\beta_0}{\beta_1}(1-e^{-\beta_1\rho z^\prime}) \cr 
& +\ln\bigl(E_\tau^i/{E_0}\bigr)e^{-\beta_1\rho z^\prime}
\Bigr]\, {E_0}\cr
&\cr
& \beta = \beta_0+\beta_1\ln(E/{E_0})\ .
}
\end{equation}
The quantity $z^\prime$ is the distance traveled by the tau after
its production. None of the choices for $\beta$ is entirely satisfactory.
Case (III) is the best approximation to the energy curves
of Fig. \ref{fig:evszcomp}. 
For case (I), 
\begin{equation}
\beta= 0.85\times 10^{-6}\ {\rm cm}^2/{\rm g}
\end{equation}
works moderately well for $E_\nu=10^{10}$ GeV, but poorly for higher
and lower energies. We reproduce the histograms of the stochastic
results from Fig. \ref{fig:evsz}
showing the last 20 km in Fig. \ref{fig:evszcomp} together with
parameterizations of $\beta$. In all cases of the parameterized energy,
we take $E_\tau^i =0.8 E_\nu$.
The dot-dashed lines show the parameterization of case (I).

Case (II) is used by Aramo et al. in Ref. \cite{aramo} with
\begin{eqnarray}
\beta_\tau & = & 0.71\times 10^{-6} \ {\rm cm}^2/{\rm g} \\
\gamma_\tau &=& 0.35\times 10^{-18} \ {\rm cm}^2/{\rm g\, GeV}\ .
\nonumber
\end{eqnarray}
The energy parameterization in this case is shown with the dashed lines
in Fig. \ref{fig:evszcomp}.

The parameterization using a logarithmic dependence on energy for 
$\beta$ does the best at reproducing the energy behavior of the emerging
taus. This is shown with solid lines in Fig. \ref{fig:evszcomp}, with
\begin{eqnarray}
\beta_0&=&1.2\times 10^{-6}\ {\rm cm}^2/{\rm g} \\
\nonumber
\beta_1&=&0.16 \times 10^{-6}\ {\rm cm}^2/{\rm g}\\
\nonumber
E_0 &=& 10^{10}\ {\rm GeV} \ .
\end{eqnarray}
Not surprisingly, this energy dependent $\beta$ will give the
best representation of the Monte Carlo simulation results for
the emerging tau fluxes from
power law incident neutrino fluxes. We note that these values of
$\beta$ are larger than what one would extract from the
direct calculation of $\beta$ \cite{drss} as in Eq. (6). This is because 
Eq. (7) is only an approximate equality.

\section{Power law neutrino spectra}

\subsection{Results from Monte Carlo simulation}

In this section, we consider cases of incident neutrino fluxes
that depend on energy. Rather than use specific flux models,
we consider power law fluxes
$F_\nu\sim E_\nu^{-n}$, for $n=1,2,3$, to see qualitatively
and quantitatively how flux propagation is affected by
tau energy loss. 

\begin{figure}[t]
\begin{center}
\epsfig{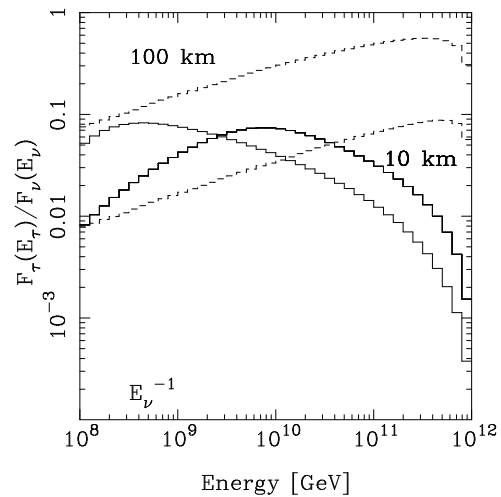}
\end{center}
\caption{Flux ratio of tau flux to incident neutrino flux as
a function of energy for 10 km and 100 km of rock, for an incident
neutrino energy scaling like $E_\nu^{-1}$. The dashed lines do not have
energy loss or decay included. The histograms include the stochastic
treatment of the neutrino interaction and
tau energy loss and decay.}
\label{fig:em1}
\end{figure}

In Fig. \ref{fig:em1},
we show the results of the stochastic propagation of
the incident tau neutrino, which converts to a tau and
continues through the depth $D$ of rock, for $D=10$ km and
100 km and an incident neutrino flux scaling like
$E_\nu^{-1}$. All of the fluxes shown in this section have a
sharp cutoff at $E_\nu=10^{12}$ GeV. The two upper dashed
curves show the
outgoing tau flux to the incoming neutrino flux neglecting tau
energy loss and decay. Because of the larger column depth
for 100 km of rock, the ratio of the tau to neutrino flux is
ten times larger than for 10 km. The flux ratio increases 
with energy to scale
with the neutrino cross section.

Our stochastic propagation 
including energy loss and decay is 
shown by the histograms in Fig. \ref{fig:em1}. Energy loss is
much stronger for the 100 km depth, so the ratio of outgoing
tau flux to incident neutrino flux is smaller for 100 km than
10 km for $E>10^9$ GeV, despite the increased number of targets for the
neutrinos. A qualitative discussion of the results for Fig.
\ref{fig:em1} is in Sec. IV.B.

\begin{figure}[t]
\begin{center}
\epsfig{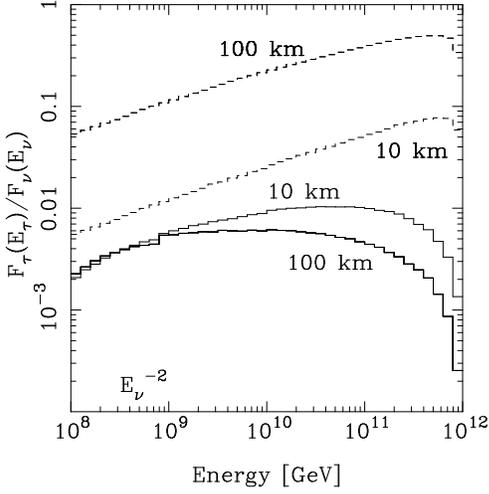}
\end{center}
\caption{Flux ratio of tau flux to incident neutrino flux as
a function of energy for 10 km and 100 km of rock, for an incident
neutrino energy scaling like $E_\nu^{-2}$. The dashed histograms 
do not have
energy loss or decay included. The histograms include the stochastic
treatment of the neutrino interaction and
tau energy loss and decay.}
\label{fig:em2}
\end{figure}

\begin{figure}[t]
\begin{center}
\epsfig{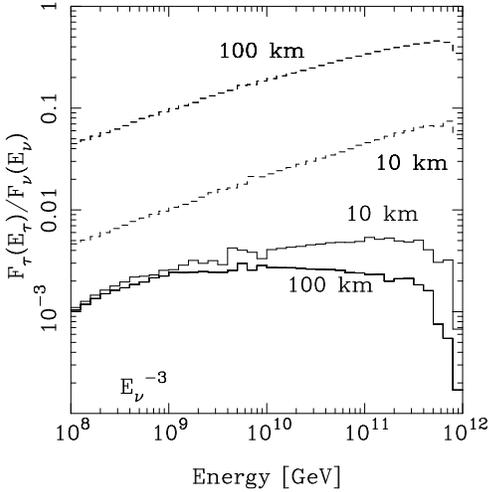}
\end{center}
\caption{Flux ratio of tau flux to incident neutrino flux as
a function of energy for 10 km and 100 km of rock, for an incident
neutrino energy scaling like $E_\nu^{-3}$. The dashed lines do not have
energy loss or decay included. The histograms include the stochastic
treatment of the neutrino interaction and
tau energy loss and decay.}
\label{fig:em3}
\end{figure}

Figs. \ref{fig:em2} and \ref{fig:em3} show the ratio of the outgoing
tau flux to the incident tau neutrino flux for incident neutrino
fluxes scaling as $E_\nu^{-2}$ and $E_\nu^{-3}$, respectively.
The flux ratios including decay and energy
loss at $E=10^8$  GeV for these
steeper fluxes can be understood based only on the lifetime,
since energy loss is not overwhelming at that energy and there is
little pile-up of taus from neutrino interactions at higher
energies, followed by tau energy loss.
The dashed lines have a factor of $D$ to account for the column
depth of nucleon targets in the neutrino interaction probability.
When decays are included, the relevant distance at $E=10^8$ GeV
is $d=E_\tau c\tau/m_\tau$, so the neutrino interaction probability
is proportional to $d$ rather than $D$. Neutrino attenuation in
10 km and 100 km of rock at $E_\nu\sim 10^8$ GeV is small, so the
flux ratios in Figs. \ref{fig:em2} and 
\ref{fig:em3}  are the same for the two depths.
At higher energies, attenuation and energy loss play a bigger role.

Our aim here is to compare the stochastic propagation to
analytic approaches. In the next section we outline the
analytic evaluation of the emerging tau flux and
make quantitative comparisons with the histograms in Fig.
\ref{fig:em1}, \ref{fig:em2} and \ref{fig:em3}.

\subsection{Analytic approximations}

The analytic approximation to the emerging tau flux as a function
of tau energy depends on a number of components. 
We review the discussion of Ref. \cite{feng} here, expanded
to include two energy dependences of $\beta$, a linear dependence \cite{aramo}
and a logarithmic one \cite{fargionboth}.

The incident flux
$F_\nu(E_\nu,0)$ is attenuated along its trajectory a distance
$z$ through the
rock with an approximate factor
\begin{equation}
F_\nu(E_\nu,z)\simeq \exp \Bigl[-z\sigma_{tot}(E_\nu)\rho N_A\Bigr]
F_\nu(E_\nu,0)\ .
\end{equation}
In our evaluation here, we are assuming a constant density
(standard rock). Neutrino regeneration\cite{halzen,
regeneration,bottai,bmss,fargionboth}.   
from neutral current
processes or tau decay is not important over these short
distances.

The probability for neutrino conversion to taus in the distance 
interval [$z,\, z+dz$] with energy in the interval
$[E_\tau^i,E_\tau^i+dE_\tau^i]$ is
\begin{equation}
P_{CC}(\nu_\tau\rightarrow \tau,E_\nu,E_\tau^i) =
{dz\,dE_\tau^i} N_A\rho \frac{d\sigma_{CC}(E_\nu,E_\tau^i)}{dE_\tau^i}\ .
\end{equation}
As discussed above, we take $E_\tau^i=0.8 E_\nu$:
\begin{eqnarray}
\frac{d\sigma_{CC}}{dE_\tau^i} &\simeq&\sigma_{CC}(E_\nu)\delta
\bigl(E_\tau^i-(1-\langle y\rangle )E_\nu \bigr)\\
\langle y\rangle &\simeq & 0.2\ .
\end{eqnarray}
Writing the differential cross section this way ensures that
the integral over tau energy yields the charged-current  (CC) cross 
section.

In the infinite lifetime, no energy loss ($\tau\rightarrow \infty$, 
$\beta\rightarrow 0$) limit, the emerging tau flux 
given incident tau neutrinos is 
\begin{eqnarray}
& F_\tau &(E_\tau^i)_{\tau\rightarrow \infty ,\beta\rightarrow 0}=
\int_0^D dz\int
dE_\nu   N_A\rho \frac{d\sigma_{CC}(E_\nu,E_\tau^i)}{dE_\tau^i}\nonumber \\ 
&\times &
\exp \Bigl[-z\sigma_{tot}(E_\nu)\rho N_A\Bigr]
F_\nu(E_\nu,0)\nonumber\\
&\equiv& \int_0^D dz\ \frac{dF_\tau^0(E_\tau^i,z)}{dz}
\end{eqnarray}
for rock depth $D$.

In the case where the energy loss is treated analytically,
the lifetime and energy loss are accounted by a factor
$P_{surv}(E_\tau^i,E_\tau,D-z)$ as the tau travels over the
remaining distance from production point $z$ to emerge from 
the rock after distance $D$ with outgoing energy $E_\tau$. 
In fact, $E_\tau^i$, $E_\tau$ and
$D-z$ are not all independent, as shown in Eq. (\ref{eq:cases}).
The survival probability as a function of energy
comes from the solution to the
equation
\begin{equation}
\frac{dP_{surv}}{dz}  =  -\frac{P_{surv}}{c\tau E_\tau /m_\tau}
\end{equation}
together with the approximate $dE_\tau/dz$ formula
of Eq. (\ref{eq:dedx2}), leading to
\begin{equation}
\frac{dP_{surv}}{dE_\tau}\simeq \frac{m_\tau}{c\tau\beta\rho E_\tau^2}
P_{surv}\ .
\label{eq:surv}
\end{equation}
The solution depends on the
energy dependence of $\beta$. For constant $\beta$, the solution is
\begin{equation}
{P_{surv}}(E_\tau,E_\tau^i) = \exp\Biggl[-\frac{ m_\tau}{c\tau\beta\rho}\Biggl(
\frac{1}{E_\tau}-\frac{1}{E_\tau^i}\Biggr)\Biggr]\quad
{\rm case\ (I)}\ .
\end{equation}
The emerging tau flux is 
\begin{eqnarray}
F_\tau(E_\tau) & = & \int_0^D dz \frac{dF^0(E_\tau^{i\prime},z)}{dz}
P_{surv}(E_\tau,E_\tau^{i\prime})\nonumber\\ 
&\times& \delta\Bigl(
E_\tau^{i\prime}-E_\tau^{i}(E_\tau,D-z)\Bigr)
\frac{dE_\tau^{i\prime}}{dE_\tau} dE_\tau^{i\prime}\nonumber\\
 & = & \int_0^D dz \frac{dF^0(E_\tau^{i\prime},z)}{dz}
P_{surv}(E_\tau,E_\tau^{i\prime})\nonumber\\ 
&\times& \delta\Bigl(
E_\tau-E_\tau(E_\tau^{i\prime},D-z)\Bigr)
dE_\tau^{i\prime}\ .
\label{eq:master}
\end{eqnarray}
The delta function explicitly enforces Eq. (\ref{eq:cases}) rewritten with
$E_\tau^i$ in terms of $E_\tau$ and $D-z$.
A factor of $dE_\tau^{i\prime}/{dE_\tau}$ accounts for the fact that
$F_\tau(E_\tau )$ represents the number of $\tau$'s per tau energy
interval, say for $\Delta E_\tau=E_2-E_1$. 
Electromagnetic energy loss means that one must sample
a much larger interval of initial tau energies, $\Delta E_\tau^i=
E_2\exp(\beta \rho D)-E_1$.  Combining this factor with the
delta function leads to the factor of
$\delta (E_\tau - E_\tau^{i\prime}\exp(-\beta\rho (D-z)))$ that
appears in the literature\cite{feng} for tau neutrino
production of taus for a constant beta:
\begin{eqnarray}
&&F_\tau(E_\tau)  = \int_0^D dz \frac{dF^0(E_\tau^{i\prime},z)}{dz}
P_{surv}(E_\tau,E_\tau^{i\prime})\nonumber\\ 
&\times& \delta\Bigl(
E_\tau-E_\tau^{i\prime}\exp\bigl(-\beta\rho(D-z)\bigr)\Bigr)
 dE_\tau^{i\prime}\ .
\label{eq:masterconst}
\end{eqnarray}

In the $\beta\rightarrow 0$ limit, the emerging tau flux is
the expected result,
\begin{equation}
F_\tau(E_\tau)  = \int_0^D dz \frac{dF^0(E_\tau^{i\prime},z)}{dz}
\exp\Bigl({-\frac{m_\tau (D-z)}{c\tau E_\tau}}\Bigr)\ .
\end{equation}

Continuing with case (I) for non-zero $\beta$, explicit substitutions give
\begin{eqnarray}
\nonumber
F_\tau(E_\tau)  &= &  \int_0^D dz \int dE_\nu \exp\Bigl({-z\sigma_{tot}(E_\nu)\rho N_A}\Bigr) F_\nu(E_\nu,0)\\ \nonumber
& \times & N_A \rho{\sigma_{CC}(E_\nu)}\delta(E_\tau -0.8 E_\nu\exp 
(-\beta\rho (D-z)))\\ 
&\times & \exp
\Biggl[-\frac{m_\tau}{c\tau\beta\rho }\Biggl( \frac{1}{E_\tau}
-\frac{e^{-\beta\rho (D-z)}}{E_\tau}\Biggr)\Biggr]\ .
\label{eq:ftaubc}
\end{eqnarray}

\begin{figure}[t]
\begin{center}
\epsfig{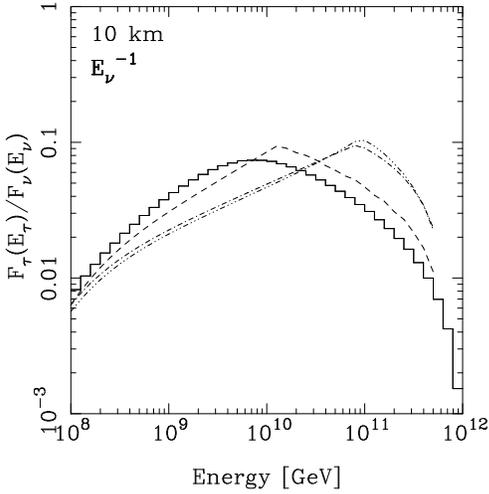}
\end{center}
\caption{Flux ratio of tau flux to incident neutrino flux as
a function of energy for 10 km of rock, for an incident
neutrino energy scaling like $E_\nu^{-1}$. 
The histograms include the stochastic
treatment of the neutrino interaction and
tau energy loss and decay. The analytic approximations for
cases (I, dot-dashed), (II, dot-dot-dot-dashed) and (III, dashed) are
also shown.}
\label{fig:hcomp1}
\end{figure}

A useful limit to understand our results in Fig. \ref{fig:em1}
is the case of constant $\beta$ in the long lifetime limit.
For simplicity, we set $\langle y \rangle = 0$ so that
$E_\tau^i=E_\nu$ \cite{feng}. If we also assume $D\ll L_{\rm int}$ where
$L_{\rm int}$ is the neutrino interaction length, we can simplify
Eq. (\ref{eq:ftaubc}) to
\begin{eqnarray}
F_\tau(E_\tau)  &= &  \int_0^D \frac{dz}{L_{\rm int}^{CC}} \int dE_\nu 
F_\nu(E_\nu,0)\delta(E_\tau - E_\nu e^ 
{-\beta\rho (D-z)})\nonumber\\
& = & \frac{1}{E_\tau \beta\rho \, L_{\rm int}^{CC}}\int_{E_{min}}^{E_{max}}
dE_\nu F_\nu(E_\nu,0), 
\label{eq:tinf}
\end{eqnarray}
for $L_{int}$ approximately constant.
The energy integral runs
from $E_{min}=E_\tau$ to $E_{max}=\min (E_\tau\exp (\beta
\rho D), 10^{12}$ GeV). At low energies and short
distances, (e.g., $E_\tau \sim 10^8$ GeV and $D=10$ km),
the maximum energy is $E_\tau \exp(\beta\rho D)$. For $F_\nu\sim
E_\nu^{-1}$, $F_\tau\sim (D/L_{\rm int}^{CC})\cdot E_\tau^{-1}$,
while for $F_\nu\sim E_\nu^{-n}$ for $n>1$,
$F_\tau\sim 1/((n-1)\beta\rho L_{\rm int}^{CC})\cdot E_\tau^{-n}$.
This means that even with strong energy loss effects, for an $E_\nu^{-1}$
flux in the appropriate energy regime, the total distance $D$ determines the
tau flux rather than the characteristic scale of the energy loss,
$1/\beta \rho$.  In the actual case of the $\tau$, the tau decay length
$\gamma c\tau$ is comparable to the energy loss scale
$1/\beta\rho$ at $E=10^8$ GeV. For the $E_\nu^{-1}$ flux, $D$
is still the dominant scale, although the details including
both energy loss and decay as well as energy dependent cross sections
have an effect, as seen in Fig. \ref{fig:em1}. Figs. \ref{fig:em2} and
\ref{fig:em3} show that this is not the case for $E_\nu^{-2}$ and
$E_\nu^{-3}$.

\begin{figure}[t]
\begin{center}
\epsfig{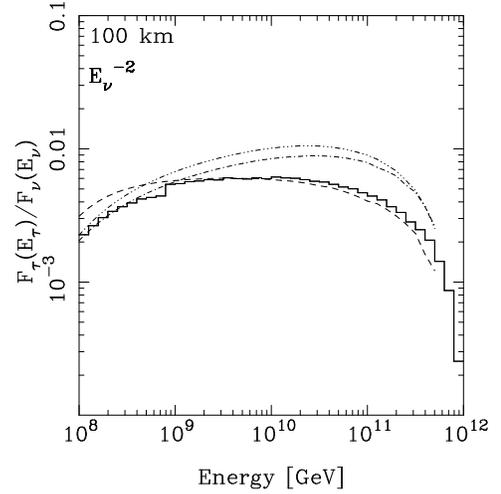}
\end{center}
\caption{Flux ratio of tau flux to incident neutrino flux as
a function of energy for 100 km of rock, for an incident
neutrino energy scaling like $E_\nu^{-2}$. 
The histograms include the stochastic
treatment of the neutrino interaction and
tau energy loss and decay. The analytic approximations for
cases (I, dot-dashed), (II, dot-dot-dot-dashed) and (III, dashed) are
also shown.}
\label{fig:hcomp2}
\end{figure}

Eq. (\ref{eq:master}) is the master equation for all three cases
discussed here for the three parameterizations of $\beta$.
The argument of the delta function comes from Eq. (\ref{eq:cases})
for the specific cases I-III.
The survival probabilities come from the solution to Eq. (\ref{eq:surv})
for the appropriate energy dependent $\beta$. This gives (see Ref.
\cite{aramo} for case (II))
\begin{eqnarray}
{P_{surv}}(E_\tau,E_\tau^i) &=& \Biggl(
\frac{E_\tau^i(\beta_\tau+\gamma_\tau E_\tau)}{E_\tau(\beta_\tau+\gamma_\tau E_\tau^i)}
\Biggr)^{\frac{m_\tau\gamma_\tau}{c\tau\beta_\tau^2\rho}}\nonumber \\
&\times &
\exp\Biggl[-\frac{ m_\tau}{c\tau\beta_\tau\rho}\Biggl(
\frac{1}{E_\tau}-\frac{1}{E_\tau^i}\Biggr)\Biggr]\quad
{\rm case\ (II)}\nonumber \\ 
{P_{surv}}(E_\tau,E_\tau^i) &=& \exp \Biggl(
\frac{m_\tau\beta_1}{c\tau\rho\beta_0^2}\Biggl[\frac{1}{E_\tau}(1+\ln
\Bigl(\frac{E_\tau}{E_0}\Bigr)\nonumber \\
&-&\frac{1}{E_\tau^i}\Bigl(1+\ln \frac{E_\tau^i}{E_0}\Bigr)\Biggr]\Biggr)
\nonumber \\
&\times &
\exp\Biggl[-\frac{ m_\tau}{c\tau\beta_0\rho}\Biggl(
\frac{1}{E_\tau}-\frac{1}{E_\tau^i}\Biggr)\Biggr]\
{\rm case\ (III)}\,.\nonumber \\
& & \label{eq:pmaster}
\end{eqnarray}
In case (III), an expansion in $\beta_1/\beta_0$ has been
made in Eq. (\ref{eq:surv}). 
Numerically, we have kept terms through $\beta_1^3/\beta_0^3$,
however in Eq. (\ref{eq:pmaster}), we show the result
where only the first term has been kept.

\begin{figure}[t]
\begin{center}
\epsfig{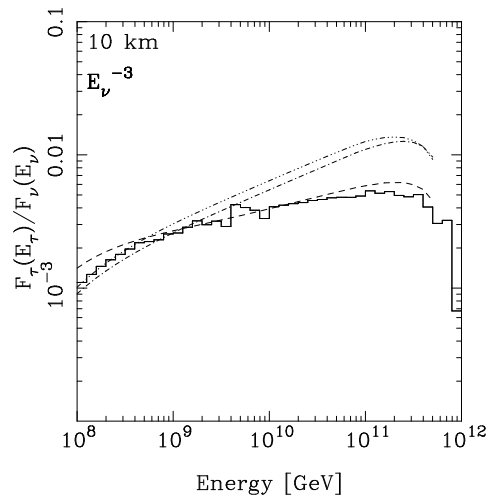}
\end{center}
\caption{Flux ratio of tau flux to incident neutrino flux as
a function of energy for 10 km of rock, for an incident
neutrino energy scaling like $E_\nu^{-3}$. 
The histograms include the stochastic
treatment of the neutrino interaction and
tau energy loss and decay. The analytic approximations for
cases (I, dot-dashed), (II, dot-dot-dot-dashed) and (III, dashed) are
also shown.}
\label{fig:hcomp3}
\end{figure}

None of the parameterizations can completely describe the Monte
Carlo results. Based on Fig. 6, it is not surprising
that the form of $\beta$ which
depends on $\log E_\tau/E_0$ works the best to describe the
outgoing tau flux. 
In Figs. \ref{fig:hcomp1}-\ref{fig:hcomp3}, we show
a comparison of the various parameterizations of the
outgoing tau flux, normalized to the incident neutrino flux.

Fig. \ref{fig:hcomp1} shows the Monte Carlo results for 10 km
of rock and an incident flux with $E_\nu^{-1}$ energy behavior
with the histogram and the case (III) parameterization with
the dashed line. The case (I) parameterization, with constant $\beta$
appears as a dot-dashed line, while the case (II) power law is shown with
a dot-dot-dot-dashed line. One sees the turnover in the flux ratio
as the change is made from $E_{max}=E_\tau \exp(\beta\rho D) $
to $E_{max}=10^{12}$ GeV for the $E_\nu^{-1}$ flux.
The parameterization
of $\beta$ as a function of $\log (E_\tau/E_0)$ (case (III)) 
gives results for
the flux ratio to within about $\pm 50$\% of the Monte Carlo result.
The other two parameterizations do not do as well, underestimating
by a factor of $\sim 2$ for $E=10^9$ GeV
and overestimating by $\sim 3$ for $E=10^{11}$ GeV.

For 100 km of rock, with the same incident flux,
the case (III) parameterization compares well
(underestimates by $\sim 15$\%) the Monte Carlo result
between $10^9-10^{11}$ GeV. By comparison, the
other two parameterizations for 100 km rock overestimate the tau
flux by a factor of $1.3-1.6$ and $1.8-2$ at energies $10^{10}$ and
$10^{11}$ GeV respectively.

For the $E_\nu^{-2}$ incident neutrino flux, the $\log E$ parameterization
of $\beta$ does well for 100 km as seen in 
Fig. \ref{fig:hcomp2}. Again, the dot-dashed
and dot-dot-dot-dashed lines show cases (I) and (II) respectively.
For incident neutrino fluxes scaling with
$E_\nu^{-3}$, the 10 km result is well represented by
the $\log E$ parameterization, as shown in Fig. \ref{fig:hcomp3}.
The overestimate of the flux for cases I and II occurs because
energy loss is underestimated at high energies with these parameterizations
of $\beta$.

\section{Discussion}

We have performed a one-dimensional Monte Carlo simulation
of $\nu_\tau$ conversion to $\tau$ and $\tau$ propagation
including electromagnetic energy loss over distances of 10-100 km
in rock. Analytic approximations
are useful tools in exploring the
possibility of detecting a variety of predictions for incident neutrino
fluxes \cite{feng}. 
Using mono-energetic sources of tau neutrinos, we modeled
the tau energy loss with an energy loss parameter $\beta$ depending
on $\log(E_\tau/E_0)$. In an analytic approximation for the
outgoing tau flux, this model for energy loss does better at representing
the Monte Carlo results than two other choices for $\beta$:
a constant and a $\beta$ increasing linearly with tau energy.
Our improved
parameterization of $\beta$ 
will help refine theoretical predictions of tau neutrino induced
events in neutrino telescopes.

\begin{figure}[t]
\begin{center}
\epsfig{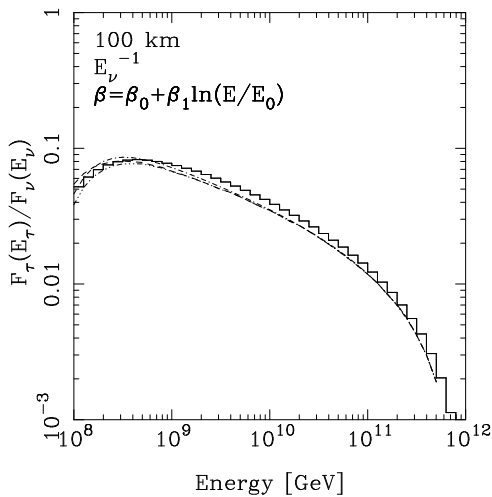}
\end{center}
\caption{Flux ratio of tau flux to incident neutrino flux as
a function of energy for 100 km of rock, for an incident
neutrino energy scaling like $E_\nu^{-1}$. 
The histogram includes the stochastic
treatment of the neutrino interaction and
tau energy loss and decay. The analytic approximations using
Eq. (28) with $\beta=\beta_0+\beta_1\ln(E_\tau/E_0)$ (dot-dashed), 
using $E_\tau(E_\tau^i,D-z)$ for case III but a simplified $P_{surv}$
(triple dot-dashed) and the full case III result 
(dashed) 
also shown.}
\label{fig:final}
\end{figure}

It is not difficult to numerically evaluate the
integrals of Eq. (\ref{eq:master}) including the more complicated 
logarithmic dependence on energy of $\beta$ (case (III)) 
for $P_{surv}$ and
$E_\tau$, nevertheless, it is interesting to see the impact of
the modifications from that energy dependence. Although formally
inconsistent, one could simply use the flux calculation for constant
$\beta$, Eq. (\ref{eq:masterconst}), and substitute $\beta=\beta_0
+\beta_1\ln (E_\tau/E_0)$. This is shown in Fig. \ref{fig:final},
for $D=100$ km and $E_\nu^{-1}$,
with the dot-dashed line. 
As in Figs.
\ref{fig:hcomp1}-\ref{fig:hcomp3}, the dashed line comes from
the full case III (logarithmic dependence)  
result, and the histogram represents the Monte Carlo result.
The dot-dashed line is nearly identical to the complete case III evaluation.
The triple dot-dashed line uses the full 
expression for the energy dependence on distance from Eq. (\ref{eq:cases}),
but it uses the simplified form of the survival probability: $P_{surv}
=\exp[-m_\tau/(c\tau\beta\rho)\cdot 
(E_\tau^{-1}-E_\tau^{i\, -1})]$. This does as well 
in representing the Monte Carlo result.
Of the fluxes and rock depths that we have discussed here,
Fig. \ref{fig:final} is representative of the
three curves incorporating to varying degrees the
logarithmic energy dependence of $\beta$. 

In summary, our Monte Carlo results for tau neutrino interaction
and tau propagation through rock suggest that the tau electromagnetic
energy loss parameter be parameterized by $\beta = \beta_0+\beta_1
\log(E/E_0)$ (case III), where the parameters $\beta_0,\ \beta_1$ and $E_0$ 
appear in Eq. (16). Using the analytic formula of Ref. \cite{feng},
modified to account for the logarithmic energy dependence,  a reasonable
agreement with the Monte Carlo results is obtained. Using
the formulae of Ref. \cite{feng} for constant $\beta$, with the small change of
taking $\langle y\rangle=0.2$ but substituting $\beta(E_\tau)$,
also is in reasonable agreement with the Monte Carlo result.

\acknowledgments

We thank I. Sarcevic and D. Seckel who were collaborators in our
initial evaluation of tau electromagnetic energy loss, and
T. Gaisser, K. Gayley, Y. Meurice and V.G.J. Rodgers.
This work is funded in part by the U.S. Department of Energy
contract DE-FG02-91ER40664.


\begin{thebibliography}{99}

\bibitem{solar}
S.~N.~Ahmed {\it et al.}  [SNO Collaboration],
Phys.\ Rev.\ Lett.\  {\bf 92}, 181301 (2004).

\bibitem{atm}
  Y.~Ashie {\it et al.}  [Super-Kamiokande Collaboration],
  Phys.\ Rev.\ Lett.\  {\bf 93}, 101801 (2004).

\bibitem{mix}
  M.~C.~Gonzalez-Garcia and M.~Maltoni,
 hep-ph/0406056.

\bibitem{halzen}
  F.~Halzen and D.~Saltzberg,
  Phys.\ Rev.\ Lett.\  {\bf 81}, 4305 (1998).

\bibitem{regeneration}
  S.~I.~Dutta, M.~H.~Reno and I.~Sarcevic,
  Phys.\ Rev.\ D {\bf 62}, 123001 (2000).

\bibitem{bottai}
 F.~Becattini and S.~Bottai,
Proceedings of the International Cosmic Ray Conference,
Salt Lake City 1999, Vol. 2, pp. 249-252 and   
Astropart.\ Phys.\  {\bf 15}, 323 (2001);
  S.~Bottai and S.~Giurgola,
  Astropart.\ Phys.\  {\bf 18}, 539 (2003).

\bibitem{bmss}
  E.~Bugaev, T.~Montaruli, Y.~Shlepin and I.~Sokalski,
  Astropart.\ Phys.\  {\bf 21}, 491 (2004).

\bibitem{fargionboth}
  D.~Fargion,
  Astrophys.\ J.\  {\bf 570}, 909 (2002); astro-ph/9704205.

\bibitem{yoshida}
  S.~Yoshida, R.~Ishibashi and H.~Miyamoto,
  Phys.\ Rev.\ D {\bf 69}, 103004 (2004).

\bibitem{reya}
  K.~Giesel, J.~H.~Jureit and E.~Reya,
  Astropart.\ Phys.\  {\bf 20}, 335 (2003).


\bibitem{feng}
  J.~L.~Feng, P.~Fisher, F.~Wilczek and T.~M.~Yu,
  Phys.\ Rev.\ Lett.\  {\bf 88}, 161102 (2002).

\bibitem{aramo}
  C.~Aramo, A.~Insolia, 
  A.~Leonardi, G.~Miele, L.~Perrone, O.~Pisanti and D.~V.~Semikoz,
  Astropart.\ Phys.\  {\bf 23}, 65 (2005).

\bibitem{convertfar}
  D.~Fargion,
  astro-ph/0412582,
  astro-ph/0106239,
  hep-ph/0206010;
 D.~Fargion, P.~G.~De Sanctis Lucentini and M.~De Santis,
  Astrophys.\ J.\  {\bf 613}, 1285 (2004) and astro-ph/0501033;

\bibitem{bertou}
  X.~Bertou, P.~Billoir, O.~Deligny, C.~Lachaud and A.~Letessier-Selvon,
  Astropart.\ Phys.\  {\bf 17}, 183 (2002).

\bibitem{athar}
  H.~Athar, G.~Parente and E.~Zas,
  Phys.\ Rev.\ D {\bf 62}, 093010 (2000).


\bibitem{jones}
 J.~Jones, I.~Mocioiu, M.~H.~Reno and I.~Sarcevic,
  Phys.\ Rev.\ D {\bf 69}, 033004 (2004).
  
\bibitem{yeh}
  P.~Yeh {\it et al.}  [NuTel Collaboration],
  Mod.\ Phys.\ Lett.\ A {\bf 19}, 1117 (2004).

\bibitem{cao}
Z. Cao, M. A. Huang, P. Sokolsky and Y. Hu,
astro-ph/0411677.


\bibitem{huang}
  M.~A.~Huang, J.~J.~Tseng and G.~L.~Lin,
{\it Proceedings of 28th International Cosmic Ray Conferences (ICRC 2003), 
 Tsukuba, Japan, 31 Jul - 7 Aug 2003} pp. 1427-1430.

\bibitem{models}
See, e.g.,   F.~Halzen and D.~Hooper,
  Rept.\ Prog.\ Phys.\  {\bf 65}, 1025 (2002) and references therein.

\bibitem{drss}
  S.~I.~Dutta, M.~H.~Reno, I.~Sarcevic and D.~Seckel,
  Phys.\ Rev.\ D {\bf 63}, 094020 (2001).

\bibitem{gqrs95}
  R.~Gandhi, C.~Quigg, M.~H.~Reno and I.~Sarcevic,
  Astropart.\ Phys.\  {\bf 5}, 81 (1996);
  G.~M.~Frichter, D.~W.~McKay and J.~P.~Ralston,
  Phys.\ Rev.\ Lett.\  {\bf 74}, 1508 (1995)
  [Erratum-ibid.\  {\bf 77}, 4107 (1996)].

\bibitem{gqrs98}
 R.~Gandhi, C.~Quigg, M.~H.~Reno and I.~Sarcevic,
  Phys.\ Rev.\ D {\bf 58}, 093009 (1998).

\bibitem{cteq6}
J.~Pumplin, D.~R.~Stump, J.~Huston, H.~L.~Lai, P.~Nadolsky and W.~K.~Tung,
  JHEP {\bf 0207}, 012 (2002).


\bibitem{andreev}
Yu. M. Andreev, V. S. Berezinsky and A. Yu. Smirnov,
Phys. Lett. B {\bf 143}, 521 (1978).


\bibitem{ekl}
R. K. Ellis, Z. Kunszt and E. M. Levin, Nucl. Phys. B {\bf 420}, 517 (1994).

\bibitem{others}
M. Gluck, S. Kretzer and E. Reya, Astropart. Phys. {\bf 11}, 327 (1999);
J. Kwiecinski, A. Martin and A. Stasto, Phys. Rev. D {\bf 56}, 3991 (1997);
K. Kutak and J. Kwiecinski, Eur. Phys. J. C {\bf 29}, 521 (2003);
R. Fiore et al., Phys. Rev. D {\bf 68}, 093010 (2003);
M. V. T. Machado, Phys. Rev. D {\bf 70}, 053008 (2004).

\bibitem{ls}
 P.~Lipari and T.~Stanev,
  Phys.\ Rev.\ D {\bf 44}, 3543 (1991).


\bibitem{music}
  P.~Antonioli, C.~Ghetti, 
  E.~V.~Korolkova, V.~A.~Kudryavtsev and G.~Sartorelli,
  Astropart.\ Phys.\  {\bf 7}, 357 (1997);
  I.~A.~Sokalski, E.~V.~Bugaev and S.~I.~Klimushin,
  Phys.\ Rev.\ D {\bf 64}, 074015 (2001).


\bibitem{ionization}
  R.~M.~Sternheimer, M.~J.~Berger and S.~M.~Seltzer,
  Atom.\ Data Nucl.\ Data Tabl.\  {\bf 30}, 261 (1984).


\bibitem{brem}
A. A. Petrukhin and V. V. Shestakov, Can. J. Phys. {\bf 46},
S377 (1981).

\bibitem{pair}
  S.~R.~Kelner, R.~P.~Kokoulin and A.~A.~Petrukhin,
  Phys.\ Atom.\ Nucl.\  {\bf 60}, 576 (1997)
  [Yad.\ Fiz.\  {\bf 60}, 657 (1997)].

\bibitem{nuc}
Based on a parameterization of the HERA data by
H. Abramowicz and A. Levy, hep-ph/9712415.

\bibitem{lkv}
  W.~Lohmann, R.~Kopp and R.~Voss,
CERN-85-03.





\end{thebibliography}
\end{document}